# Beware of the water: Hidden hydrogenation of perovskite membranes made by the water-soluble sacrificial layer method

*Umair Saeed,[1,2*] Felip Sandiumenge,[3] Kumara Cordero-Edwards,[1] Jessica Padilla-Pantoja[1], José Manuel Caicedo Roque,[1] David Pesquera,[1] José Santiso,[1] Gustau Catalan[1,4†]*

[1]Catalan Institute of Nanoscience and Nanotechnology (ICN2), CSIC and BIST, Campus UAB, Bellaterra, Barcelona, 08193 Spain.
[2]Universitat Autònoma de Barcelona, Plaça Cívica, Bellaterra, Barcelona, 08193 Spain.
[3]Institute of Materials Science of Barcelona (ICMAB-CSIC), Campus UAB, Bellaterra, Barcelona, 08193 Spain.
[4]ICREA - Institució Catalana de Recerca i Estudis Avançats, Barcelona, 08010 Spain.

**ABSTRACT**. The fabrication of perovskite oxide free-standing films (membranes) by lift-off methods using water-soluble sacrificial layers such as $Sr_3Al_2O_6$ is appealing because of the new possibilities that these membranes present over conventional epitaxial films. However, little is known about how the fabrication process itself, and in particular the exposure to water during the etching step, affects their properties. Here, we investigate how membranes of two perovskite archetypes, antiferroelectric $PbZrO_3$ and paraelectric $SrTiO_3$, are affected by this fabrication process. Using Raman spectroscopy and X-ray diffraction, we find evidence that hydrogen penetrates the perovskite structure. Concomitant with this protonation, the functional properties also change, and both materials display ferroelectric-like behavior that is absent in bulk or in hydrogen-free films at room temperature. We also find that thermal annealing can be used to expel the hydrogen from the membranes, which henceforth recover their bulk-like properties.

## I. INTRODUCTION.

The field of functional oxide thin films has been revolutionized by the development of water-soluble sacrificial layers that allow the growth of fully oriented perovskite thin films that can then be released from their substrate, thereby eliminating the mechanical clamping. [1,2] These de-clamped ¨free-standing¨ films or ¨membranes¨(used interchangeably in this article) often display superior functional properties (faster switching speeds, lower losses, more efficient energy storage) as compared to the epitaxial ones in ferroelectric (FE) oxides such as $BaTiO_3$, [3,4] $BiFeO_3$ [5,6] and in antiferroelectrics (AFE); [7–10] while also enabling multiple layer stackings and thus partake in twistronics research. [11,12] Their large mechanical flexibility also enables strain-mediated domain control [13,14] or enhanced photostriction. [15]

However, while the mechanical benefits of releasing from the substrate are becoming clear, it is not known whether the etching process used to achieve such release has any chemical side-effects that might be unwittingly affecting the properties of the perovskite membranes. In particular, the typically used sacrificial layer, $Sr_3Al_2O_6$ (SAO), dissolves in water thanks to the hydrolysis of its $Al_6O_{18}^{18-}$ rings. [1] This reaction results in the formation of $H^+$ and $OH^-$ ions. Since these species are being formed exactly at the interface of the functional oxide film, there is a clear risk that they may penetrate the perovskite film.

Meanwhile, and independently, there are decades-worth of research looking at how the properties of functional oxides change when hydrogen is deliberately introduced into their lattice electrochemically or by annealing in hydrogen-containing environments. [16,17] Such studies, performed in ceramics or epitaxial films, consistently report that protonation can induce lattice expansion (strains as large as 13% have been reported [18]) and affect functional properties, in systems as varied as $NdNiO_3$, [18] $SrCoO_{2.5}$, [19,20] $SrRuO_3$ [21] and $CaRuO_3$. [22] In $PbZr_xTi_{1-x}O_3$ (PZT), first principles calculations predict polarization to increase at low H-content and decrease at higher contents, [23] and protonation has been shown to increase leakage currents in PZT capacitors. [24]

Therefore, we know on the one hand that ions are generated in the fabrication of membranes, and on the other hand that, if hydrogen is intercalated into a perovskite lattice, it changes its properties. In this context, an inevitable question emerges: does any of the hydrogen generated during the fabrication process of the perovskite membranes get inside them? And, if so, how does this unwitting protonation change their properties? Last but not least: is it possible to mitigate or even remove this hydrogen by-product from the lattice? The present article seeks to answer these questions.

*Contact author: umair.saeed@icn2.cat
†Contact author: gustau.catalan@icn2.cat

For the present study, we choose two perovskite oxides as case examples: $PbZrO_3$ (PZO) and $SrTiO_3$ (STO). PZO is regarded as an archetypal antiferroelectric (although its complex energy landscape includes other low energy polar phases), [25] while STO is a cubic paraelectric that can be driven into a ferroelectric state by strain. [26] We have synthesized membranes of PZO and STO via the standard SAO dissolution method, [1,4] and characterized their structure and properties. We have done this both for as-transferred membranes without any further treatment, and for thermally post-annealed ones. The results demonstrate that (i) the SAO dissolution method indeed can cause protonation of the membranes, affecting their structural and functional properties, and (ii) hydrogen can be removed from the lattice by suitable post-annealing.

## II. METHODS

ºOur thin films were grown by pulsed laser deposition on STO substrates coated with a SAO layer that was dissolved in water at room temperature (typical dissolution times were 4 hours), and the membranes were transferred onto silicon (experimental procedure- Supplementary Information S1). The sample thicknesses were, for PZO, 75, 36 and 17 nm, and, for STO, 16 nm, and we characterized them before and after annealing in vacuum (for 75 nm PZO and 17 nm PZO), in oxygen (36 nm PZO) or in air (for STO). We choose to anneal two of the PZO membranes in vacuum to make sure that any eventual effect of annealing could not be attributed to removal of oxygen vacancies.

## III. RESULTS

### *1. Lead Zirconate (PZO)*

We used EELS (Supplementary Information S2) and Raman spectroscopy to confirm the presence of hydrogen in the lattice. The Raman peak at 3730 $cm^{-1}$ is known to correspond to an O-H stretching vibration. [24,27] The intensity of this peak as a function of annealing time, for the PZO membrane of 75 nm annealed at 260°C in 0.5 mbar pressure of air, is shown in Figure 1(a). The peak decreases with increasing annealing time, consistent with removal of hydrogen from the lattice. The PZO spectra before and after annealing are shown in Figure 1(b). The peaks identified for both cases match closely with those reported in the literature for orthorhombic PZO. [28–30] The peaks are at 35, 45, 51, 55 and 133 $cm^{-1}$ for Pb-($ZrO_3$) lattice mode; 204 and 238 cm-1 for O-Zr-O bending; 287, 327 and 346 $cm^{-1}$ for $ZrO_3$ torsion; and 537 $cm^{-1}$ for Zr-O stretching. We observe peak shifts to lower frequencies after annealing, more pronounced for those at lower frequencies (35 to 130 $cm^{-1}$), but we do not see evidence for an annealing-induced change of symmetry. Afterwards, we immersed the membrane in deionized water for 18 hours, but the O-H stretching peak does not reappear (Figure 1(a)). This result suggests that the mere presence of water is not enough to cause hydrogen penetration into the perovskite; it is caused by the generation of $H^+$ and $OH^-$ ions during the SAO etching.

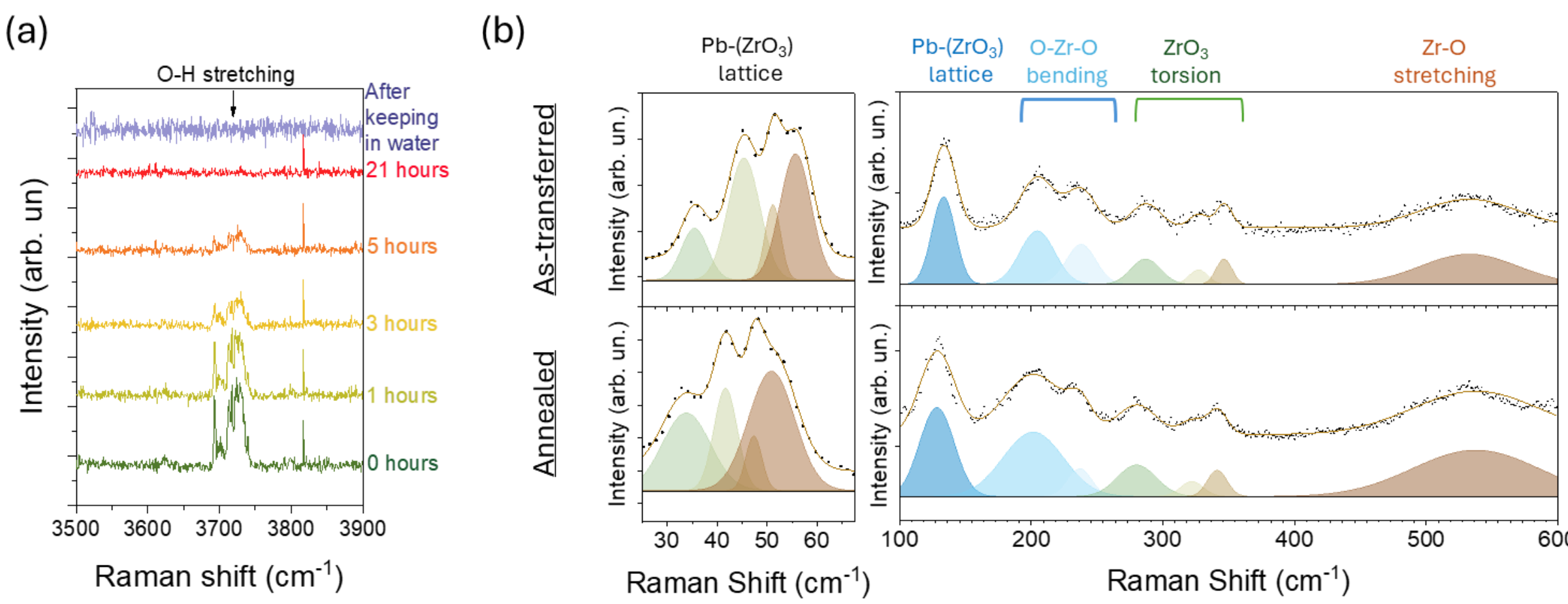

FIG 1: (a) In-situ monitoring of O-H stretching Raman peak while vacuum annealing at 260oC and later immersing in DI water, (b) Room temperature Raman spectra for as-transferred (top) and annealed (bottom) PZO membrane with the assigned modes labelled at top (the scatter dots are data points, the brown line shows the total fitting curve).

*Contact author: umair.saeed@icn2.cat

†Contact author: gustau.catalan@icn2.cat

The systematic red shift of the Raman spectra suggests a change in lattice strain upon annealing. We have therefore used X-ray diffraction (XRD) to characterize the strain state of our membranes (Figure 2). The Θ-2Θ scans of the thicker PZO film (75 nm) grown epitaxially on the SAO buffer layer indicate high crystalline quality with a predominant $(120)_o = (100)_{pc}$ orientation and minority $(001)_o = (001)_{pc}$ -oriented domains. The peak position associated to these domains is maintained after the release and transfer of the PZO film onto the silicon substrate, although the rocking curve broadens from 0.06° to 0.5°, indicating inhomogeneous strain. After annealing, the FWHM of the rocking curve slightly decreases to 0.4° (Figure 2 (a)). The more significant differences, however, are in the in-plane direction. The reciprocal space map (RSM) of the epitaxial film shows an in-plane lattice parameter of 4.13 Å, representing a -0.7% in-plane compressive strain with respect to the bulk value of 4.16 Å(Supplementary Information S3) [10, 30] while the as- transferred membrane displays an in-plane lattice expansion of 0.7% (a = 4.19 Å) with respect to the bulk. After annealing, the membrane recovers the bulk-like lattice parameter of 4.16 Å (Figure 2 (b)).

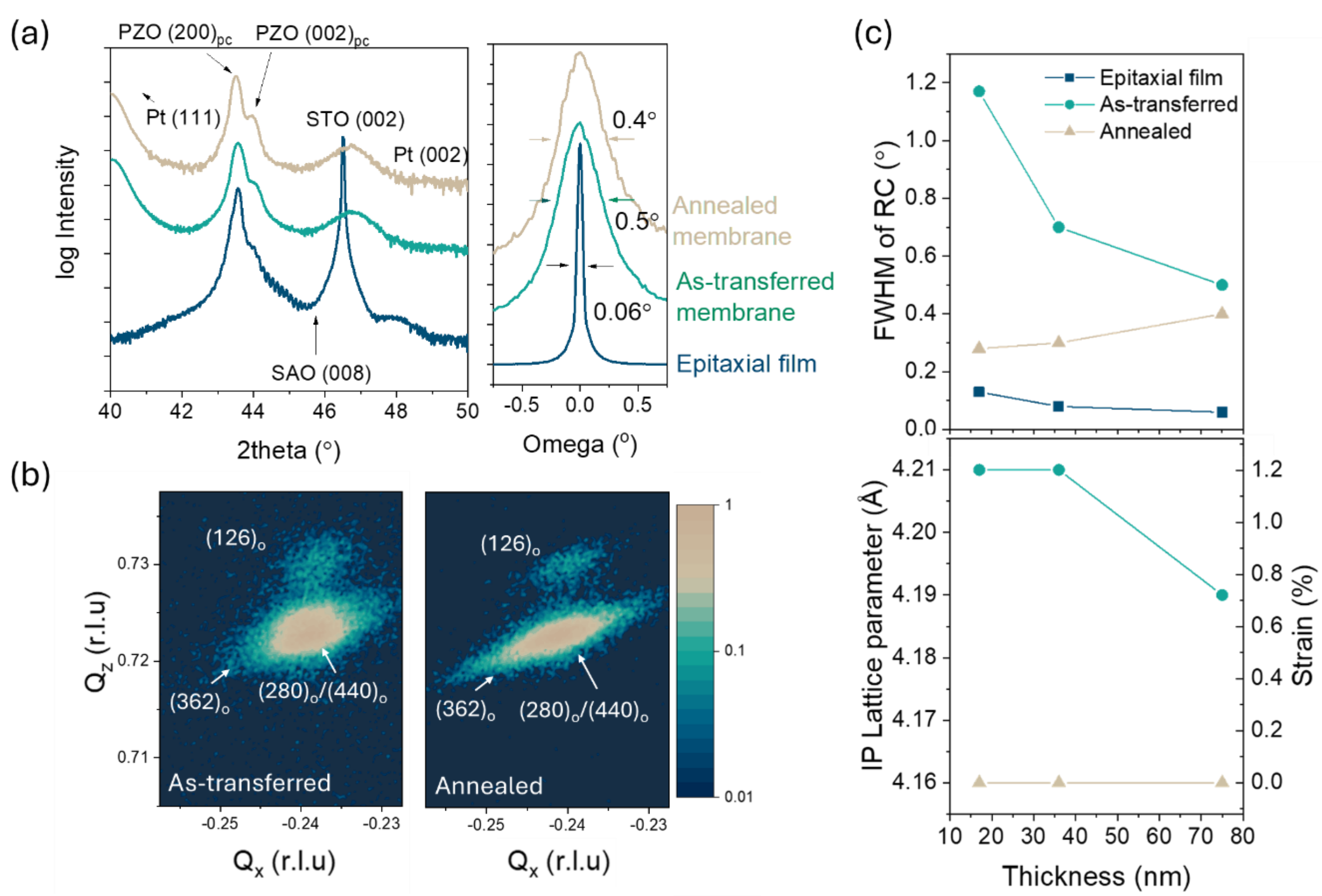

FIG 2: XRD for the PZO membranes at different stages, (a) θ-2θ XRD scans (left) and rocking curves around (002) reflection (right) for the 75 nm PZO epitaxial film, as-transferred and annealed membrane, (b) RSM around (-103)pc peak of PZO for the as-transferred (left) and annealed membrane (right), (c) Evolution of FWHM of RCs of membranes after each stage of processing and thickness dependence of in-plane lattice parameter and calculated strain evolution for the as- transferred and annealed PZO membranes (for the peak originating from (100)pc / (120)o domains).

*Contact author: umair.saeed@icn2.cat

†Contact author: gustau.catalan@icn2.cat

We attribute this lattice expansion, and the consequential volume expansion in the as-transferred membrane to the incorporation of hydrogen atoms in the lattice during the SAO etching process. It is nevertheless surprising that the out-of-plane lattice parameters at every stage of processing remain the same as bulk (4.16 Å). Other than the protonation-induced strain (in-plane expansion and inhomogeneous strain broadening of diffraction peaks), the actual lattice symmetry does not seem to change, i.e. the number and relative height of the diffraction peaks remain the same, in agreement with the Raman results.

We also characterized 36 nm and 17 nm PZO using XRD (Supplementary Information S5- S7). The in-plane lattice expansion in these thinner membranes is higher (1.2%) than for the 75 nm PZO (Figure 2(c)). This result is consistent with a finite penetration depth of the hydrogen ions, which therefore affects thinner films more than thicker ones. A depth-dependent implantation depth is also consistent with the thickness dependence of the inhomogeneous strain, indicated by the FWHM of the rocking curves.

To examine functionality, we use piezoresponse force microscopy (PFM) in dual AC resonant tracking mode (DART) [31] to characterize the voltage-induced deformation of the 75 nm membrane before and after annealing (Figure 3) and for 36 nm PZO membrane (Supplementary Information S6). We firstly scan a 3 × 3 $\mu m^2$ area applying a positive voltage to the tip (+10 V) and then a 1 × 1 $\mu m^2$ area in the center of the previous square with a negative bias (−10 V). The PFM phase and amplitude read after these poling scans show a canonical ferroelectric-like behavior in the as-transferred membrane, with clear 180° phase contrast and similar piezoelectric amplitude between both domains, and minimum amplitude at the boundary between them. The unwritten region outside of the larger square has much lower amplitude and an intermediate phase with respect to those of the written regions. This result suggests that the membrane transitions from the AFE to FE-like phase by writing and does not switch back to the AFE phase within the timescale of the measurement (4 minutes). Over time, though, the membrane does relax back to the antiferroelectric phase, as indicated by a fading contrast in phase and amplitude in the subsequent PFM scans. The behavior of the released membranes would also be consistent with an electret-type response of the implanted ions. The conclusion, in any case, is that while the ground state of the membranes is non-polar, metastable macroscopic polarization can be induced in them by a voltage.

In contrast, after annealing the membrane, applying the same poling process does not lead to any contrast in the PFM phase and amplitude. Hence, we deduce that the annealed membrane is robustly antiferroelectric and the field-induced polar phase, if it exists, is unstable and switches back before there is time to read it. To verify these conclusions, we examine the piezoresponse hysteresis loops via switching spectroscopy (SS-PFM). [32]

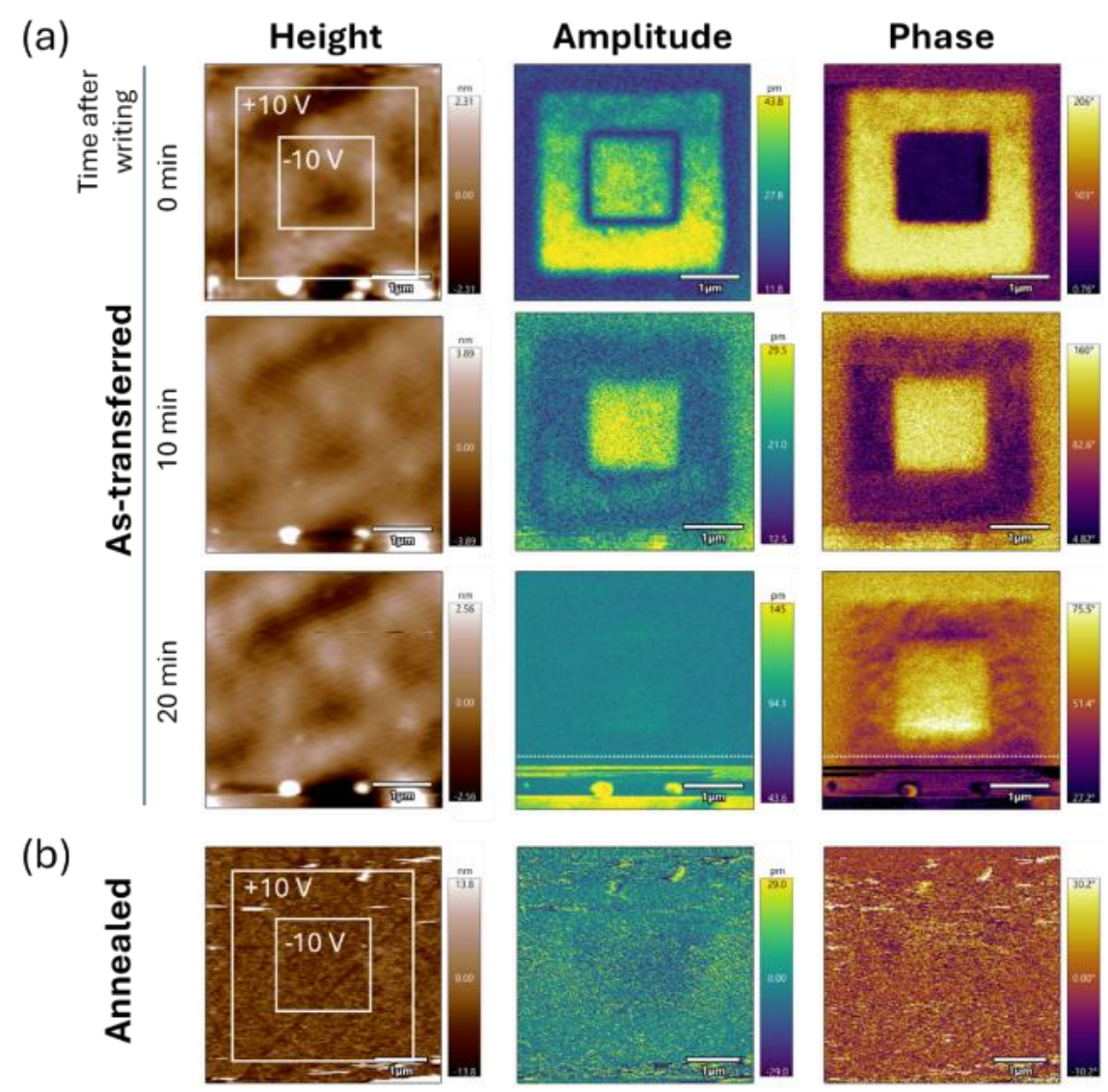

FIG 3: AFM topography, PFM amplitude and phase after writing square patterns indicated for the (a) as-transferred 75 nm PZO membrane and their time evolution (Data below dashed line in the bottom panels is unreliable due to strong crosstalk with topography), (b) annealed PZO membrane showing no polarization switching.

A pulsed triangular wave, with superimposed AC bias, has been applied through the AFM tip (Supplementary Information S1). The values for amplitude and phase were recorded at each on (during bias application) and off (after bias has been applied) stage, labelled as on-and off- loops respectively. Conventionally, the off-loops are used in ferroelectric systems, as they capture the remnant polarization of stable domains after they have been oriented by the external voltage. However, in AFE systems, polarization is expected to switch back to zero when the bias is off, hence the on- loops are needed to detect the voltage-induced polar phase. [33]
Both the on and off loops of the as-transferred membrane look ferroelectric-like (the SS-PFM within a pre-written region is shown in Supplementary Information S4), with the on-loop showing higher amplitude and lower critical field than for the off-loop (Figure 4), which is expected as on-loops can have piezoelectric contributions from non-remnant polarization. For the annealed sample, however, the on and off loops are qualitatively different. We see a sudden increase in amplitude in the on- loop at around 5 V (marked by arrows in Figure 4) accompanied by additional loops in on- phase representing an AFE- FE transition. This sudden increase in amplitude is consistent with the expected increase of piezoelectric coefficient ($d_{33}$) at the field-induced transition to the polar state, typical of antiferroelectrics. [33–35] The off- loops still show some residual hysteretic behavior, reflecting incomplete relaxation of the polarization. [10,36–39]

*Contact author: umair.saeed@icn2.cat

†Contact author: gustau.catalan@icn2.cat

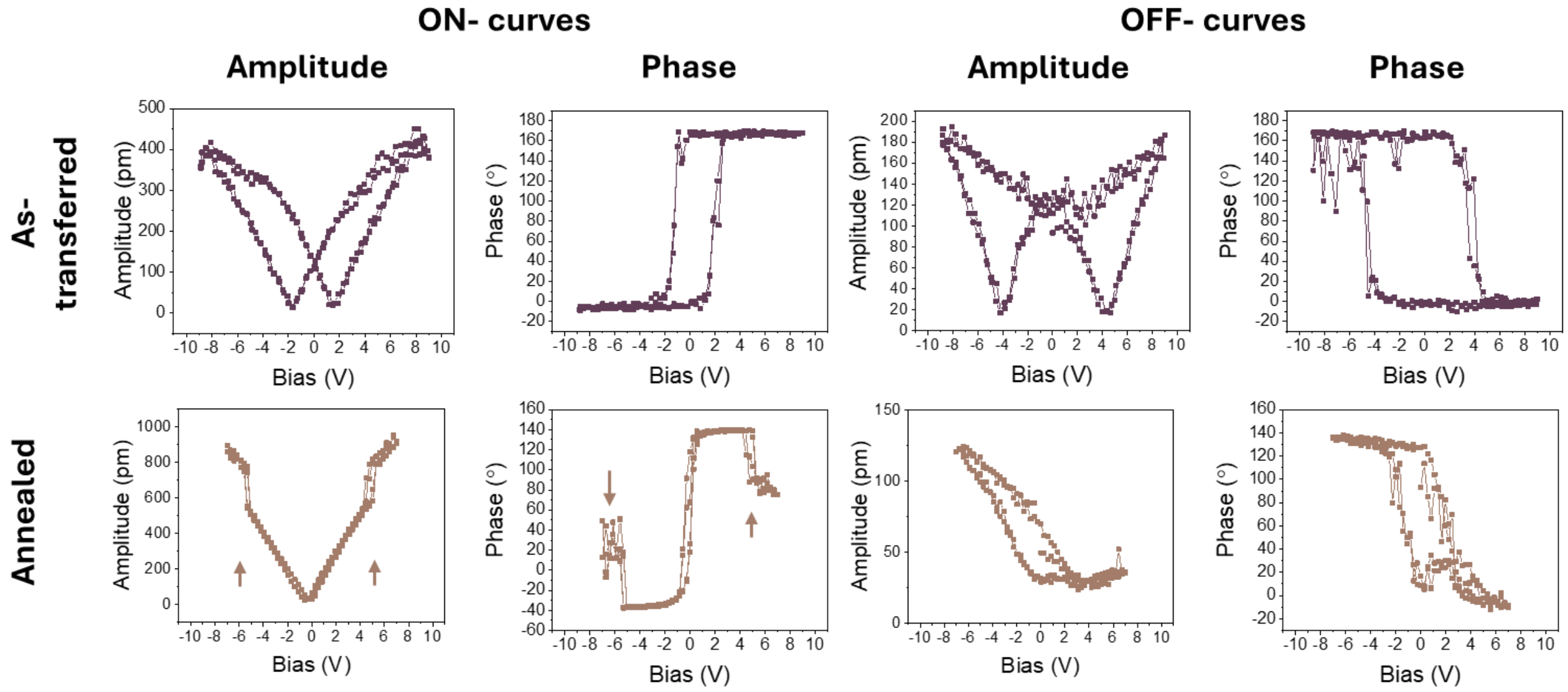

FIG 4: SS-PFM of a 75 nm PZO membrane (top panels: as-transferred, bottom panels: annealed), showing (from left to right) amplitude and phase loops in on- and off- stage.

The results thus indicate that the penetration of hydrogen into the lattice not only changes the strain, but also the functional properties, helping stabilize the field-induced polar phase of PZO; conversely, annealing gets rid of the hydrogen and standard antiferroelectric response is recovered.

### *2. Strontium Titanate (STO)*

To determine whether hydrogenation is unique to PZO or more general, we also fabricated a membrane of STO (16 nm), grown on SAO-buffered STO (001) substrate, and then transferred to Pt-coated Silicon (for PFM) and bare Silicon (for Raman spectroscopy). We have performed XRD (Figure 5 (a-b)), Raman (Figure 5 (c-d)) and SS-PFM (Figure 5 (e)) before and after annealing at 1000°C for 20 min in air, which according to the OH Raman signal (figure 5 (c)) is sufficient for bulk STO to exchange hydrogen with the atmosphere. The temperature used was much higher in STO because of the higher enthalpy of hydrogen incorporation as reported in the literature. [40]
Contrary to PZO, the presence of hydrogen does not cause significant lattice expansion in the STO membrane, showing in-plane and out-of-plane lattice parameters of 3.898 and 3.907 Å in the as-transferred membrane; and 3.903 and 3.906 Å in the annealed membrane, respectively (bulk value = 3.905 Å). The as-transferred membrane shows higher tetragonality, which in STO is often associated with ferroelectricity. Moreover, Raman spectroscopy reveals differences upon annealing (Figure 5(c) and Supplementary Information S8): the as-transferred membrane shows first-order scattering phonon modes at 153 cm$^{-1}$ (TO2 and LO1), 450 cm$^{-1}$ (TO4) and, importantly, at 798 cm$^{-1}$ (LO4). The latter is a polar phonon mode associated with the out-of-phase displacements of oxygen ions with respect to Ti. [41] First-order scattering modes are normally prohibited in STO due to its cubic centrosymmetric structure, so their presence suggests that the as-transferred STO membrane can be ferroelectric. This hypothesis is supported by SS-PFM with ferroelectric-like loops (Figure 5(e)). After annealing, the Raman spectrum no longer reveals the first-order LO4 mode, consistent with a recovery of the bulk-like paraelectric nature of the film, also supported by the lack of hysteresis in the SS-PFM.

*Contact author: umair.saeed@icn2.cat
†Contact author: gustau.catalan@icn2.cat

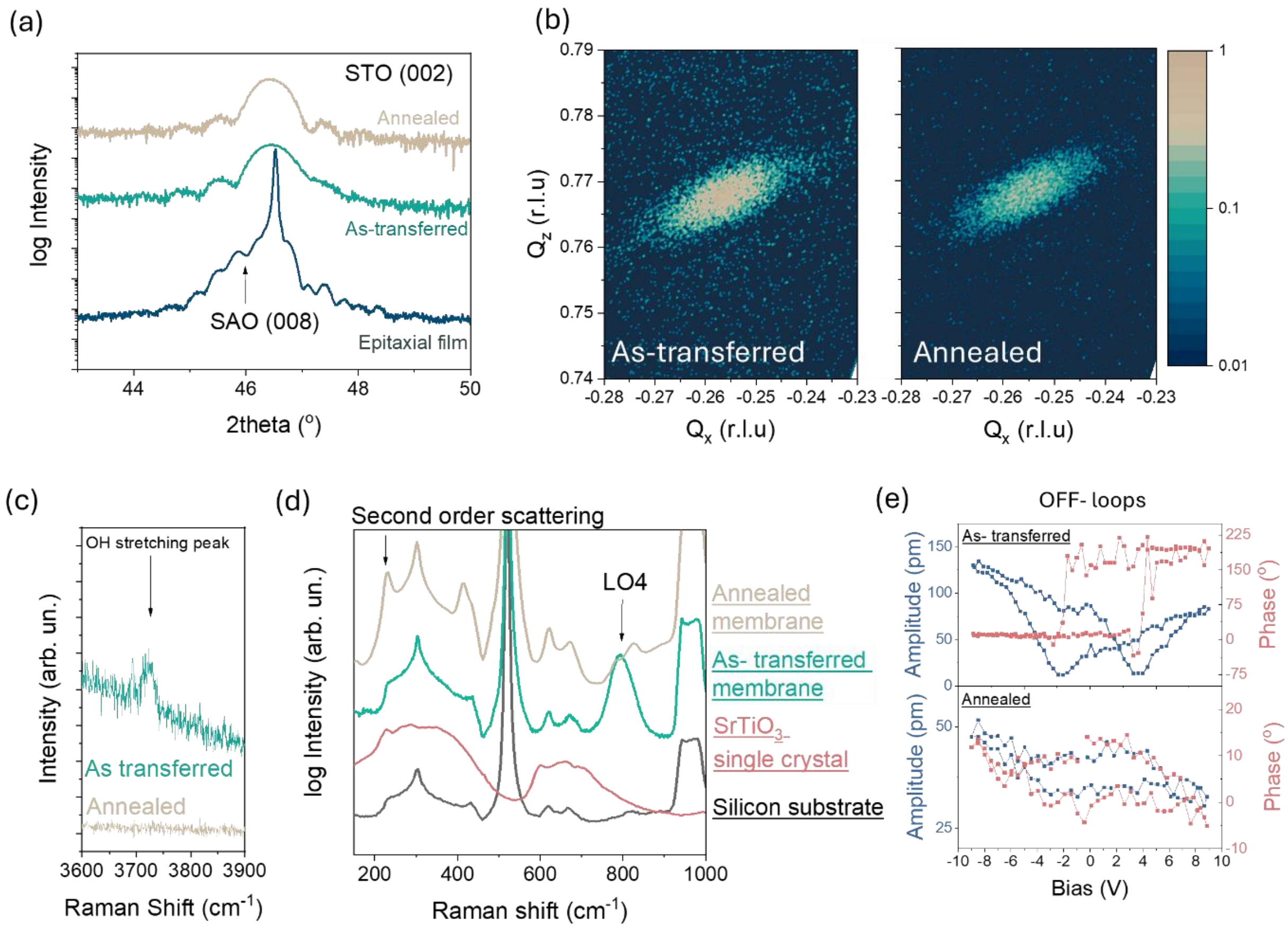

FIG 5: Characterization of STO membrane, (a) θ-2θ scans of the as-grown, as-transferred and annealed STO film, (b) RSMs around the (-103) peak of STO for the as-transferred (left) and annealed (right) membrane, (c) O-H stretching Raman peak for the as-transferred STO membrane (top) not present in the annealed membrane (bottom), (d) Raman spectra of as- transferred and annealed STO membrane with Silicon and STO single crystal spectra as references, (e) SS-PFM loops in OFF- stage for the as-transferred (top) and annealed (bottom) membrane.

## IV. DISCUSSION

The intercalation of hydrogen in the lattice of perovskite oxides is known to be possible by using acidic mediums. [42] This, in combination with enhanced activity of chemical reactions based on geometry [43] causes this protonation of membranes where the ions are provided by the reaction of SAO and water. The effects of the hydrogenation of the perovskites may be material specific. In this context, there have been several experimental and theoretical studies that focus on the role of hydrogen in the lattice of perovskite oxides. [18,19,21–24,27,44] Hydrogen is known to be able to intercalate at interstitial sites between the oxygen octahedra at varying distances depending on the symmetry of the system (octahedra tilts). [44–46] The difference in induced strain between PZO and STO at lower thicknesses might be due to PZO having a larger lattice and therefore bigger interstitial sites that facilitate the incorporation of hydrogen (PZO also has a higher electronegativity difference between the A and B atoms in the ABO3 structure than STO). [44] In $PbZrO_3$ and other perovskites, antiferroelectricity is coupled to oxygen octahedral tilts, [47–49] and such tilts decrease with tensile strain (as per Goldschmidt's tolerance factor), so the intercalation-induced lattice expansion in the plane of the film may weaken antiferroelectricity and favour ferroelectricity. While in our PZO samples, the strain does not induce a stable ferroelectric state, the PFM results do indicate a bigger propensity to polarization in the hydrogenated samples than in the annealed ones. Meanwhile, in STO, its smaller lattice parameter compared to that of PZO (3.90 Å vs 4.16 Å) complicates the interstitial percolation of $H^+$, and instead hydrogen takes, in the words of Tarun and McCluskey, a more "invisible" role replacing Sr vacancies.[40] Despite its relatively moot effect on the lattice strain, the incorporation of hydrogen in the STO membrane can still affect its symmetry and functional properties via defect dipoles that can favor and/or mimic ferroelectricity. [27,40,50–52]

*Contact author: umair.saeed@icn2.cat

†Contact author: gustau.catalan@icn2.cat

## V. CONCLUSIONS

These results highlight several lessons: (1) protonation may be a general occurrence in perovskite membranes produced by the sacrificial SAO layer method; (2) it can have different impacts on their structure (large anisotropic strain but no change of symmetry in PZO, small strain but change of symmetry in STO), and (3) despite the structural differences, the presence of hydrogen can have similar functional consequences; in our case, we observe that it favors polarization both in antiferroelectric PZO and paraelectric STO. On the plus side, annealing is seen to be effective at getting rid of the hydrogen and recovering bulk-like properties in the membranes. Therefore, post-annealing of membranes is deemed as a necessary step to obtain membranes closer to the intrinsic state of the materials under investigation.

## ACKNOWLEDGMENTS

All research at ICN2 is supported by a Severo Ochoa Grant CEX2021-001214-S. The ICMAB is funded the Severo Ochoa Centres of Excellence Programme, funded by the Spanish Research Agency (AEI, CEX2023-001263-S). U. Saeed acknowledges the funding for this project by Grant PID2019-108573GB-C21 (FOx-Me) funded by the Spanish Ministry of Science and Innovation (DOI: 10.13039/501100011033). U.S. and K.C.E also acknowledge funding by Grant N° 964931 (TSAR, DOI: 10.3030/964931) from the European Union's Horizon 2020 research and innovation program. D.P. acknowledges funding from 'la Caixa' Foundation fellowship (ID 100010434). K.C.E. and D.P. acknowledge Grant No. PID2022-140589NB-I00 funded by MCIN/AEI/10.13039/501100011033.

## SUPPLEMENTARY MATERIAL

Supplementary Information contains additional measurements and results that support the findings of this study, including structural characterization of 75 nm $PbZrO_3$ epitaxial film before etching of the sacrificial layer, 36 nm and 17 nm $PbZrO_3$ free standing membranes and Raman spectrum of $SrTiO_3$ membrane on Pt/Si substrate.

*Contact author: umair.saeed@icn2.cat
†Contact author: gustau.catalan@icn2.cat

*Contact author: umair.saeed@icn2.cat

†Contact author: gustau.catalan@icn2.cat

*Contact author: umair.saeed@icn2.cat

†Contact author: gustau.catalan@icn2.cat

*Contact author: umair.saeed@icn2.cat
†Contact author: gustau.catalan@icn2.cat

*Contact author: umair.saeed@icn2.cat
†Contact author: gustau.catalan@icn2.cat

# Supplementary Information

Beware of the water: Hidden hydrogenation of perovskite membranes made by the water-soluble sacrificial layer method

*Umair Saeed[1,2*], Felip Sandiumenge[3], Kumara Cordero-Edwards[1], Jessica Padilla-Pantoja[1], José Manuel Caicedo Roque[1], David Pesquera[1], José Santiso[1], Gustau Catalan[1,4†]*

[1]Catalan Institute of Nanoscience and Nanotechnology (ICN2), CSIC and BIST, Campus UAB, Bellaterra, Barcelona, 08193 Catalonia.

[2] Universitat Autònoma de Barcelona, Plaça Cívica, 08193 Bellaterra, Barcelona, Catalonia.

[3]Institute of Materials Science of Barcelona (ICMAB-CSIC), Campus UAB, Bellaterra, Barcelona, 08193 Catalonia.

[4]ICREA - Institució Catalana de Recerca i Estudis Avançats, Barcelona, Catalonia, 08010.

E-mail: *umair.saeed@icn2.cat
†gustau.catalan@icn2.cat

*Contact author: umair.saeed@icn2.cat

†Contact author: gustau.catalan@icn2.cat

**S1: Experimental procedures:**

To fabricate membranes, we grew 5 nm of sacrificial layer, $Sr_3Al_2O_6$ (SAO) on a $SrTiO_3$ (100) (STO) followed by functional films using pulsed laser deposition. SAO was grown with a fluence of 2.5 J/cm$^2$ using a laser frequency of 2 Hz in Oxygen partial pressure of 1 mTorr at 750˚C. $PbZrO_3$ (PZO) was grown with fluence of 1.66 J/cm$^2$, 5 Hz laser frequency and 100 mTorr oxygen partial pressure at 575˚C, whereas STO with a fluence of 1.8 J/cm$^2$ at 2 Hz laser frequency in 100 mTorr oxygen partial pressure at 750˚C. After growth, the sample was attached to a support polymer: Polydimethylsiloxane (PDMS) from the film side and then placed in water, where SAO etched away, leaving the membrane on the PDMS. The PDMS was then brought in contact with a secondary substrate, Platinum coated Silicon, heated to 60˚C and was slowly raised using a micrometer probe, transferring the membrane on the secondary substrate. The schematic is shown in **Figure FS1a** (PZO as an example)**.** Finally, the annealing of the PZO membranes was done at 260˚C in vacuum (0.5 mbar) for 75 nm and 17 nm PZO, in dynamic oxygen environment for 36 nm PZO with a flow rate of 18 sccm and at 1000˚C in air for 16 nm STO membrane.

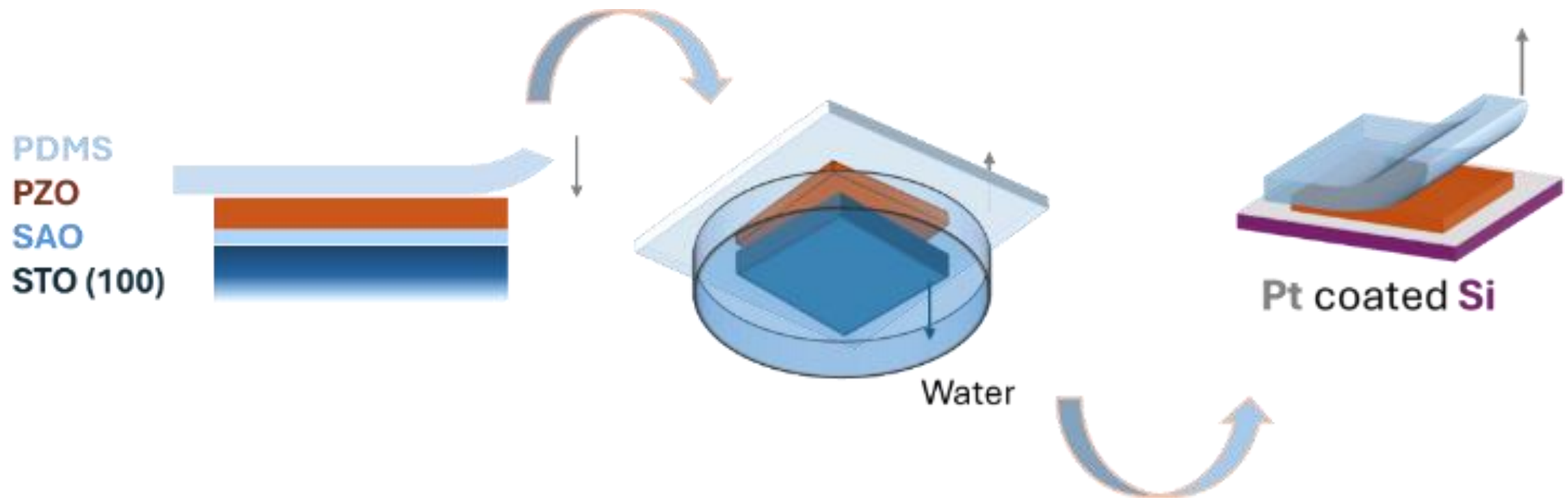

*FS 1a: Schematic of transfer of membranes to a secondary substrate.*

.

For in-situ vacuum annealing with RAMAN measurements, Nextron probe station was used to heat and cool the PZO membrane with a rate of 5˚C/min attached to the Pfeiffer MVP 015-4 diaphragm pump (final pressure of 375 mtorr (= 0.5 mbar). The XRD was measured using Panalytical X'pert Pro diffractometer (Copper K-$\alpha_1$, 1.540598 Å), using a hybrid 2-bounce primary monochromator on the incident beam side and a PIXcel line detector. Raman measurements were done using Witec RAMAN alpha300 access (Oxford instruments) using a laser of 532 nm with laser power of 1.75 mW.

PFM was done using Asylum Research MFP-3d AFM (Oxford instruments). SS-PFM loops were measured with a pulsed triangular wave at a frequency of 0.1 Hz, with the time-period of ON and OFF stage equal to 0.14 seconds (**Figure FS1b**). The whole wave is superimposed with an AC wave of 0.7 V (not shown) to measure the piezoresponse.

*Contact author: umair.saeed@icn2.cat

†Contact author: gustau.catalan@icn2.cat

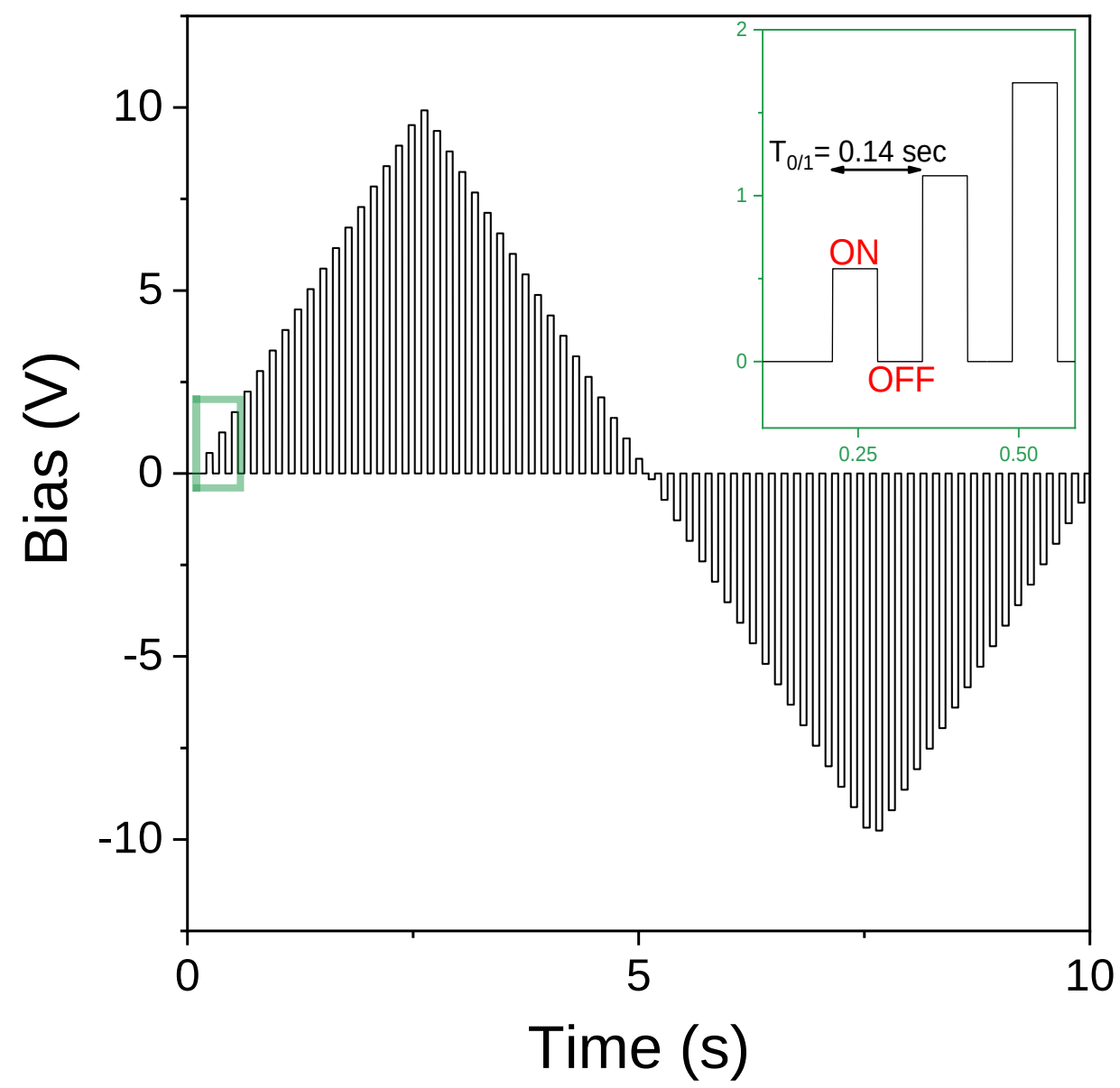

***FS 1b**: Pulsed triangular wave applied to obtain SS-PFM loops.*

*Contact author: umair.saeed@icn2.cat

†Contact author: gustau.catalan@icn2.cat

## S2: Electron energy loss spectroscopy (EELS) of as-transferred PZO membrane.

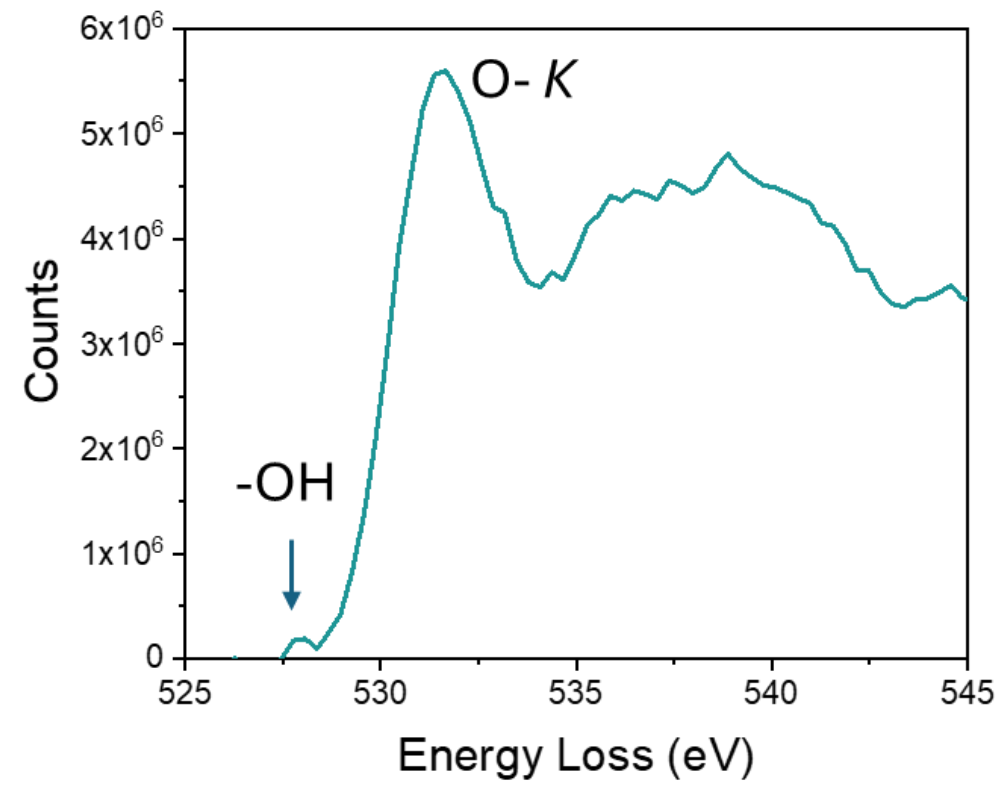

***FS 2****: EELS of as-transferred PZO membrane shows a peak at 528 eV (corresponding to -OH) showing the presence of Hydrogen bonded with lattice oxygen in the $PbZrO_3$ lattice.*

*Contact author: umair.saeed@icn2.cat

†Contact author: gustau.catalan@icn2.cat

**S3: RSM of epitaxial film of 75 nm PZO and SAO heterostructure on STO (100) substrate**

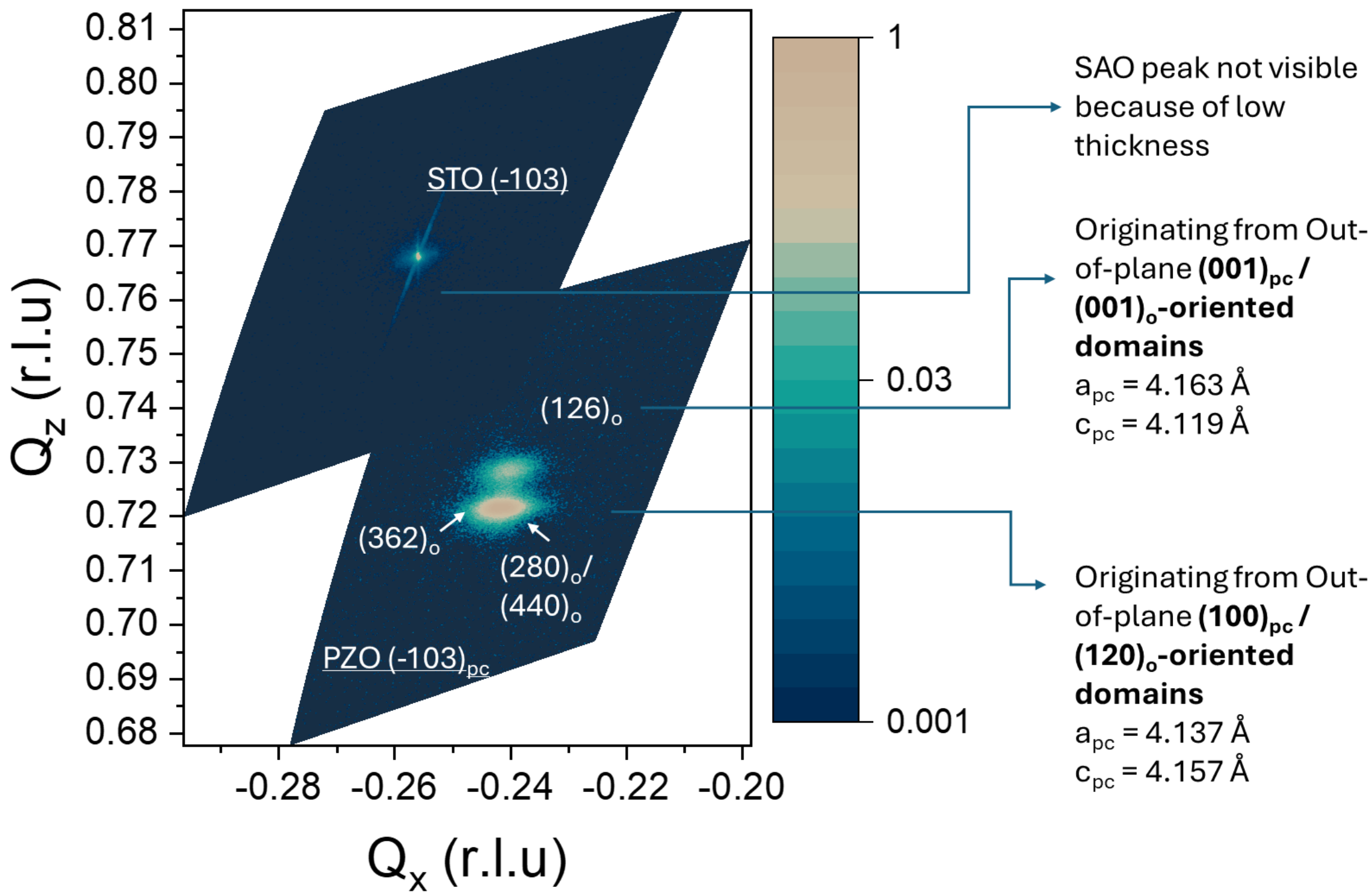

*Contact author: umair.saeed@icn2.cat

†Contact author: gustau.catalan@icn2.cat

**S4: SS-PFM in the written region of the as-transferred PZO membrane of 75 nm.**

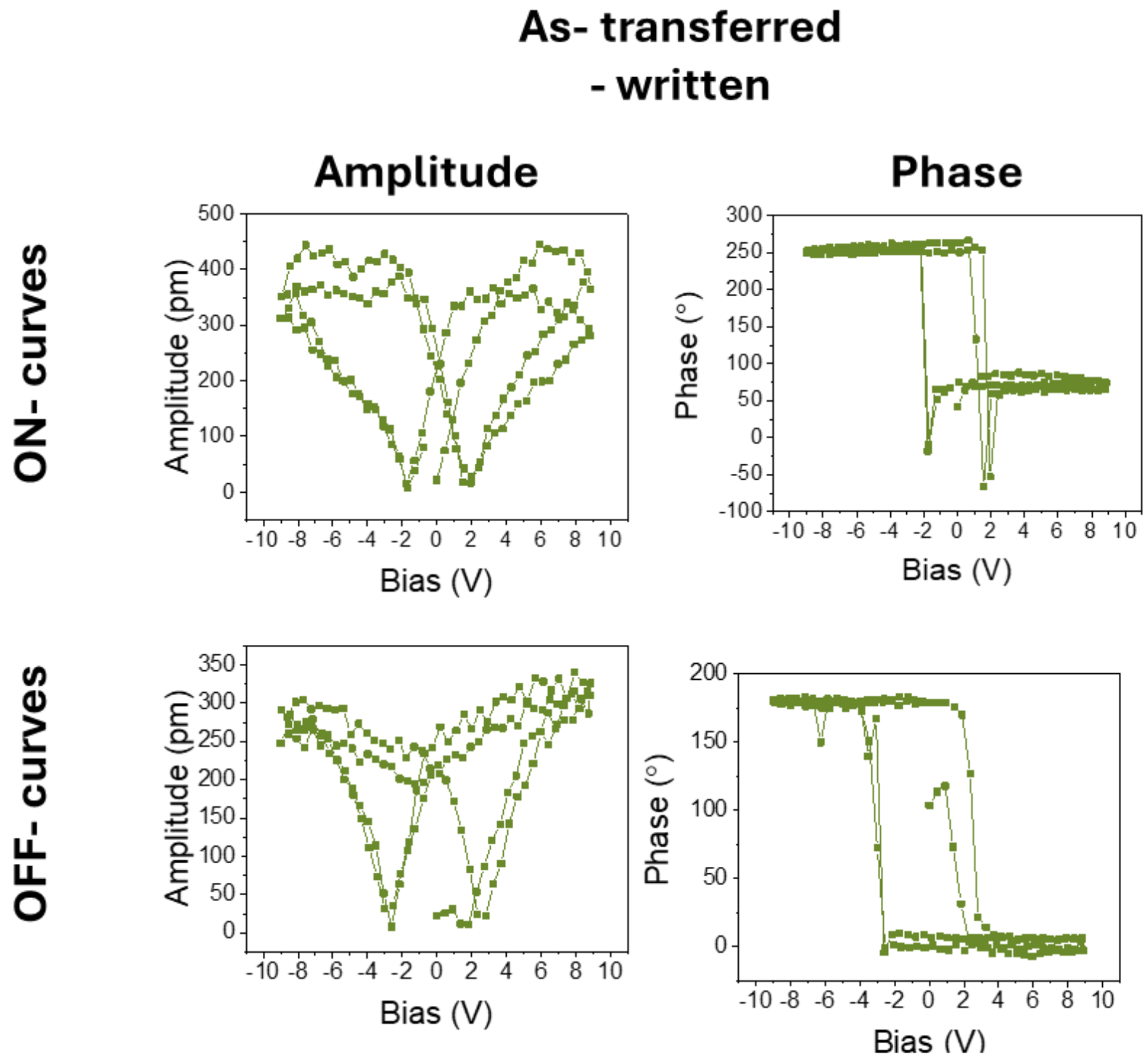

In the written area of the as-transferred PZO membrane, the loops are also ferroelectric-like. Notably, the values of both the amplitude change and coercive field in *on*- and *off*- loops are more similar than in the pristine region, suggesting an increased stability of the polarized domains -i.e. a lower presence of fast-relaxing polarization. This indicates that written regions have a more stable ferroelectric response, with a lower presence of spontaneous back-switching, than the pristine ones.

*Contact author: umair.saeed@icn2.cat
†Contact author: gustau.catalan@icn2.cat

**S5: Structural characterization of 36 nm PZO film before and after annealing in Oxygen environment.**

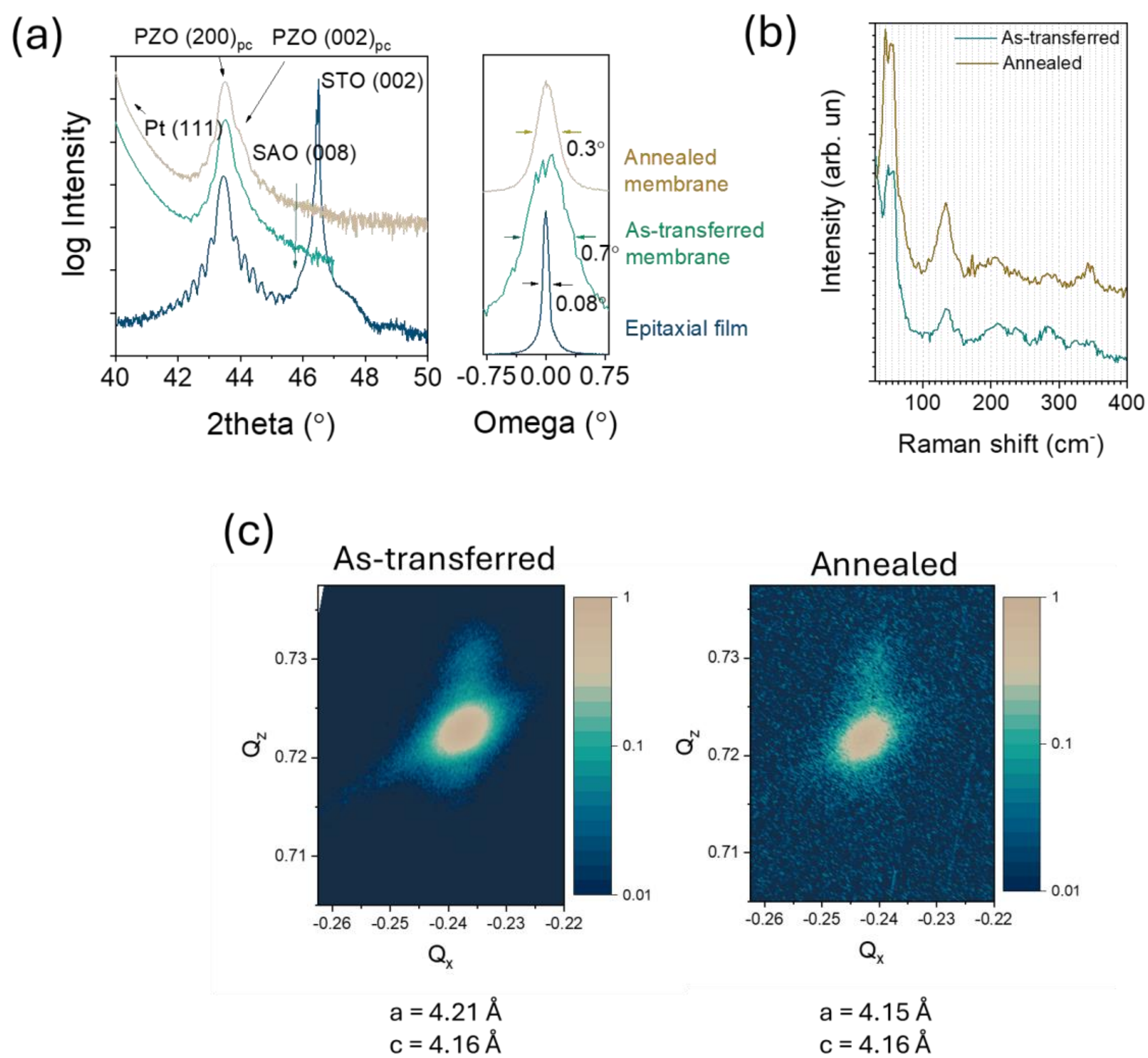

*FS 5: Results for 36 nm PZO:* ***(a)*** *2-theta scans (left) and Rocking curves (right) for PZO film at different stages,* ***(b)*** *RAMAN of 36 nm PZO before and after vacuum annealing showing red shift of peaks,* ***(c)*** *RSMs around $(-103)_{pc}$ peak of PZO for the as-transferred and annealed membrane.*

*Contact author: umair.saeed@icn2.cat

†Contact author: gustau.catalan@icn2.cat

**S6: PFM results for 36 nm PZO before and after annealing in Oxygen environment.**

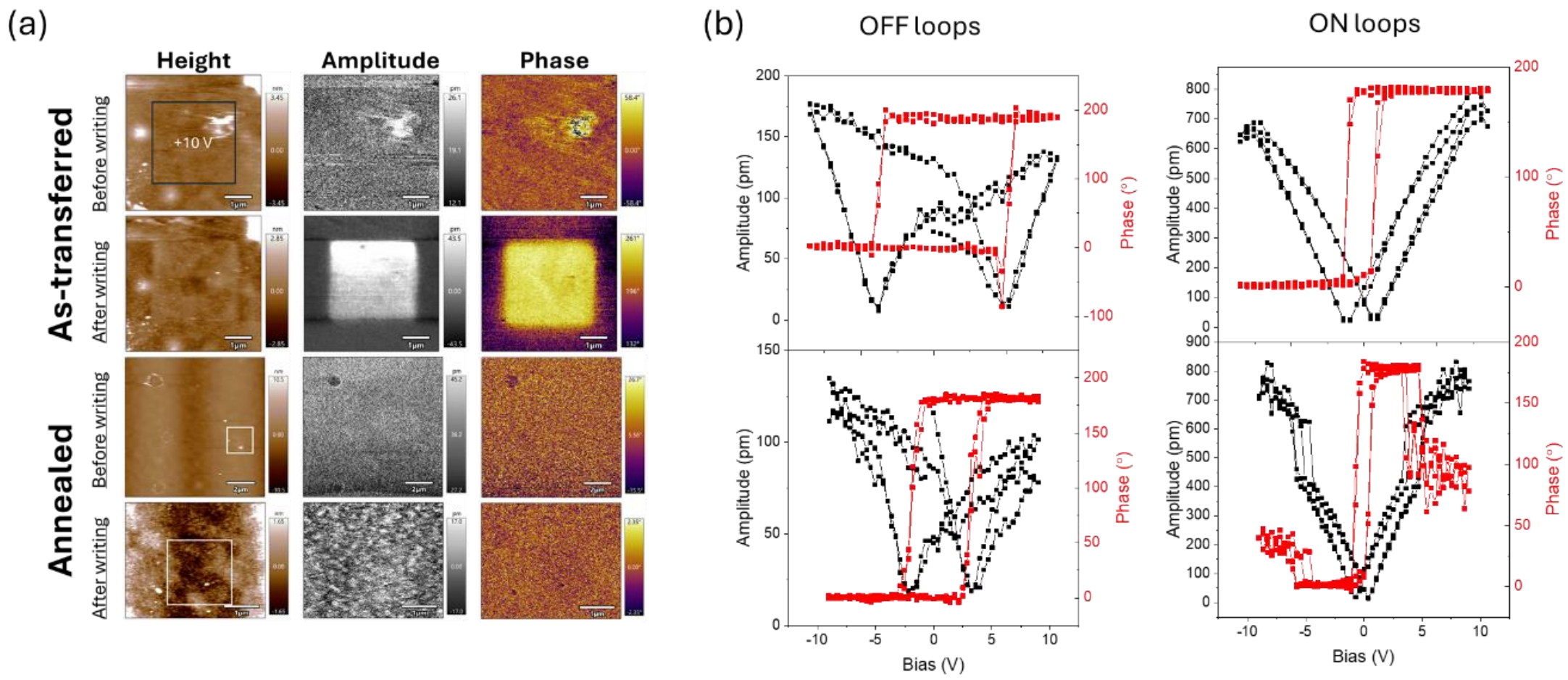

*FS 6: PFM of 36 nm PZO membrane on Pt-Si substrate, **(a)** writing with +10 V on as-transferred and vacuum annealed membrane indicated by the blue and white boxes, **(b)** SS-PFM loops on the as-transferred (top panels) and oxygen annealed membrane (bottom panels).*

The 36 nm PZO membrane annealed in oxygen shows the same behaviour as the 75 nm membrane annealed in vacuum. The *off*- loop for the annealed membrane shows a much clearer ferroelectric- type loop for this thinner membrane, which might be due to the coexistence of ferroelectric or ferrielectric phases that arise because of size dependant and surface effects. However, annealing in Oxygen takes less time (4 hours) as compared to vacuum annealing (21 hours as seen in in-situ RAMAN measurements).

*Contact author: umair.saeed@icn2.cat

†Contact author: gustau.catalan@icn2.cat

**S7: Structural characterization of 17 nm PZO membrane before and after annealing in vacuum.**

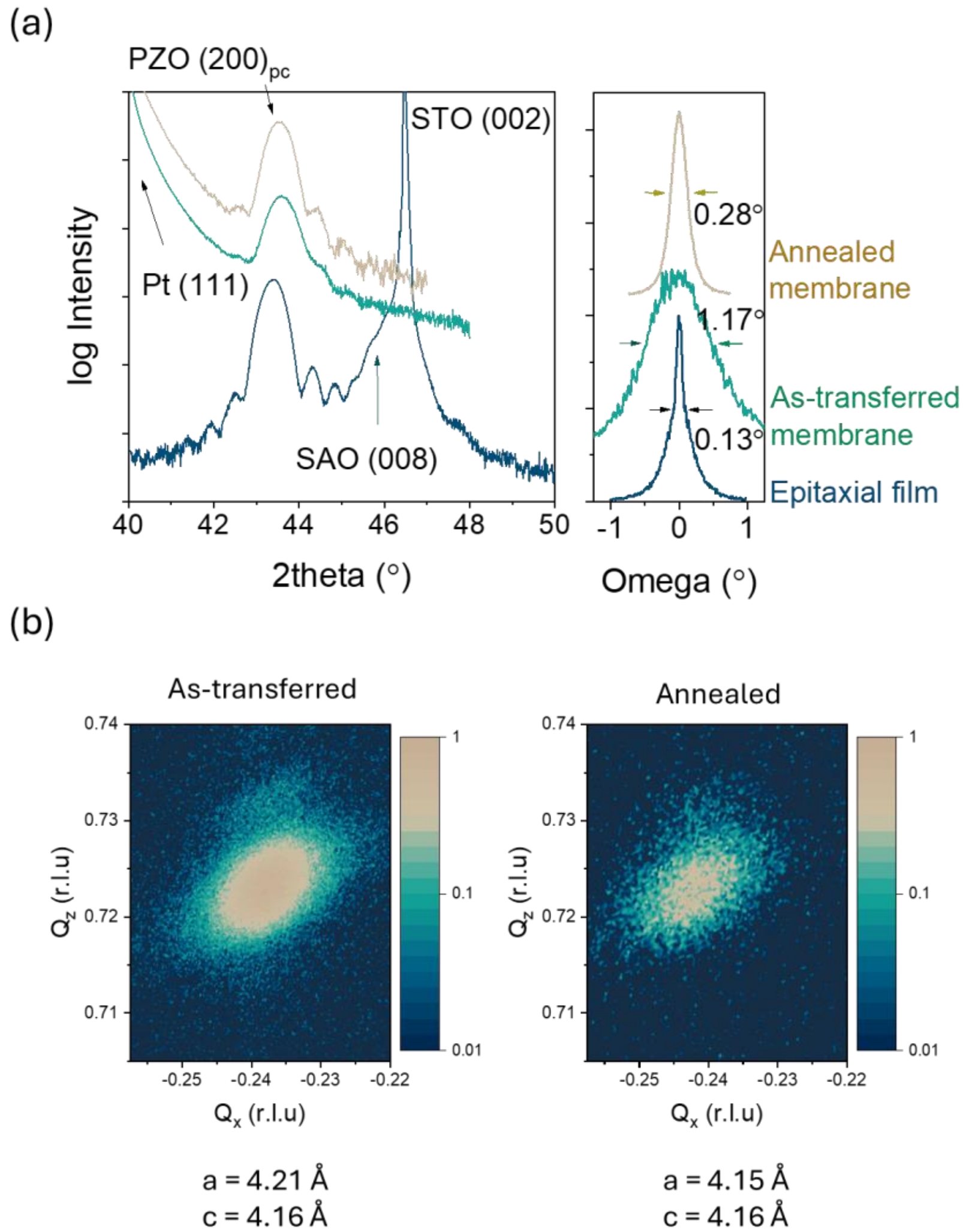

*FS 7*: ***(a)*** *2theta scans (left) and RC comparisons (right) and* ***(b)****: RSMs around $(-103)_{pc}$ peaks of 17 nm PZO for the As- transferred (left) and Annealed (right) membrane.*

*Contact author: umair.saeed@icn2.cat
†Contact author: gustau.catalan@icn2.cat

## S8: Characterization of STO membrane

**Figure FS 8(a)** shows Raman of STO membrane transferred to Pt-Si substrate, showing the first order scattering modes (TO2, LO1, TO4 and LO4) along with second order scattering peaks marked in the black squares. RAMAN on Pt-Si substrate prevents the acquisition of signal from the silicon underneath. However, the sample on Pt-Si cannot be annealed at high temperatures because of the wetting of the metal layer, exposing the silicon. Therefore, for better comparison, bare silicon substrate is used to compare the Raman spectra for the as- transferred and annealed membrane in the main text. As reference, we also include the spectra for single crystal STO and silicon. However, since most of the second order scattering peaks are overshadowed by the peaks for silicon in the annealed membrane, we also probe higher frequency peaks of STO at 1300 and 2440 $cm^{-1}$ as shown in **(b)** to make sure we are indeed measuring the STO spectra and not just the background since there are no silicon peaks in these regions.

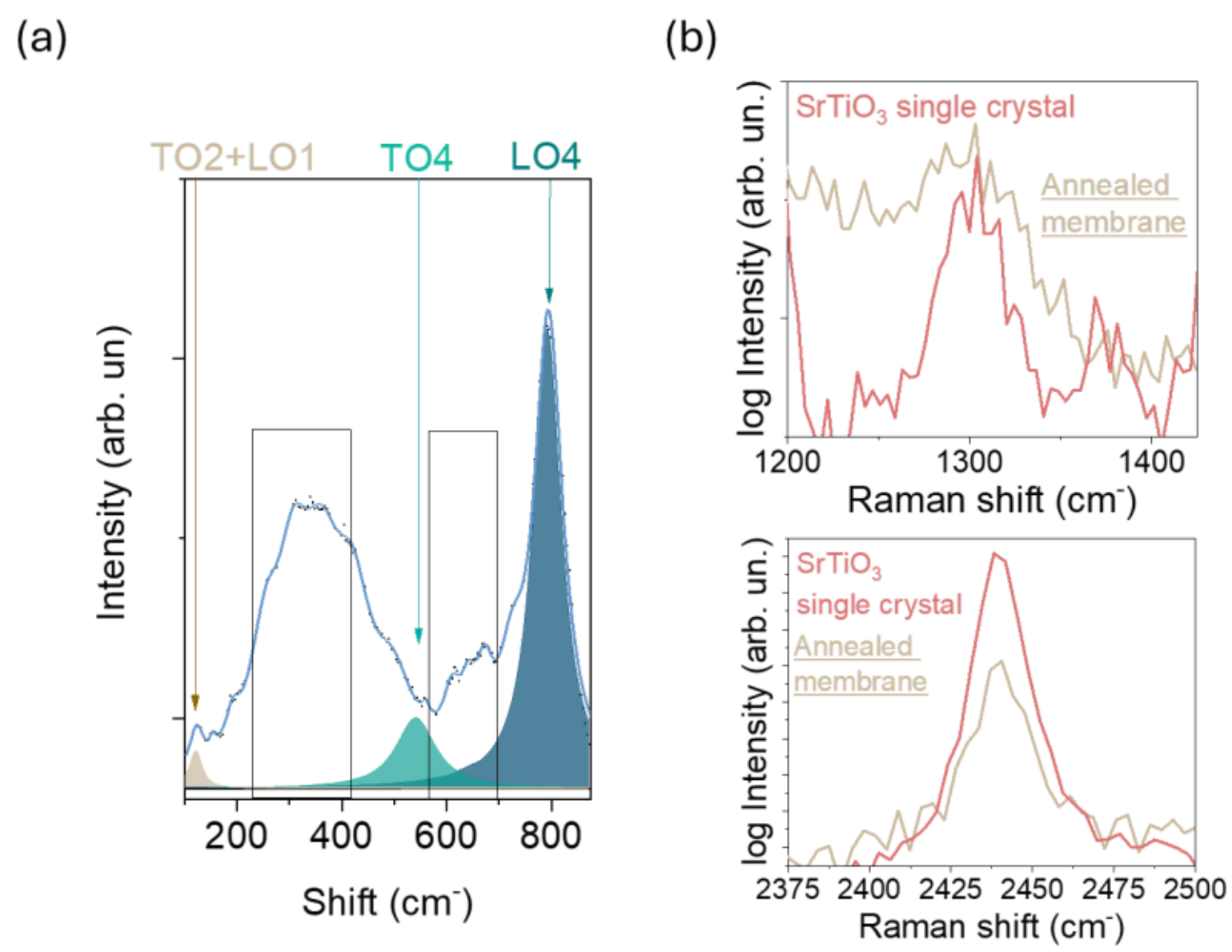

*FS 8: **(a)** Raman of STO membrane on Pt-Si before annealing, **(b)** high frequency modes of STO membrane on bare Silicon.*

*Contact author: umair.saeed@icn2.cat

†Contact author: gustau.catalan@icn2.cat